\documentclass{emulateapj} 
\usepackage{graphicx}
\usepackage{amssymb,amsfonts,amsmath,amstext,amsgen,amsopn,amsxtra,indentfirst,times}

\begin{document}

\title{Understanding Trends Associated with Clouds in Irradiated Exoplanets}
\author{Kevin Heng\altaffilmark{1}}
\author{Brice-Olivier Demory\altaffilmark{2}}
\altaffiltext{1}{University of Bern, Center for Space and Habitability, Sidlerstrasse 5, CH-3012, Bern, Switzerland.  Email: kevin.heng@csh.unibe.ch}
\altaffiltext{2}{Department of Earth, Atmospheric and Planetary Sciences, Massachusetts Institute of Technology, 77 Massachusetts Ave., Cambridge, MA 02139, U.S.A.  Email: demory@mit.edu}

\begin{abstract}
Unlike previously explored relationships between the properties of hot Jovian atmospheres, the geometric albedo and the incident stellar flux do not exhibit a clear correlation, as revealed by our re-analysis of Q0--Q14 \textit{Kepler} data.  If the albedo is primarily associated with the presence of clouds in these irradiated atmospheres, a holistic modeling approach needs to relate the following properties: the strength of stellar irradiation (and hence the strength and depth of atmospheric circulation), the geometric albedo (which controls both the fraction of starlight absorbed and the pressure level at which it is predominantly absorbed) and the properties of the embedded cloud particles (which determine the albedo).  The anticipated diversity in cloud properties renders any correlation between the geometric albedo and the stellar flux to be weak and characterized by considerable scatter.  In the limit of vertically uniform populations of scatterers and absorbers, we use an analytical model and scaling relations to relate the temperature-pressure profile of an irradiated atmosphere and the photon deposition layer and to estimate if a cloud particle will be lofted by atmospheric circulation.  We derive an analytical formula for computing the albedo spectrum in terms of the cloud properties, which we compare to the measured albedo spectrum of HD 189733b by Evans et al. (2013).  Furthermore, we show that whether an optical phase curve is flat or sinusoidal depends on whether the particles are small or large as defined by the Knudsen number.  This may be an explanation for why Kepler-7b exhibits evidence for the longitudinal variation in abundance of condensates, while Kepler-12b shows no evidence for the presence of condensates, despite the incident stellar flux being similar for both exoplanets.  We include an ``observer's cookbook" for deciphering various scenarios associated with the optical phase curve, the peak offset of the infrared phase curve and the geometric albedo.
\end{abstract}

\keywords{planets and satellites: atmospheres}

\section{Introduction}
\label{sect:intro}

\subsection{Observational Motivation}

\begin{figure}
\centering
\includegraphics[width=\columnwidth]{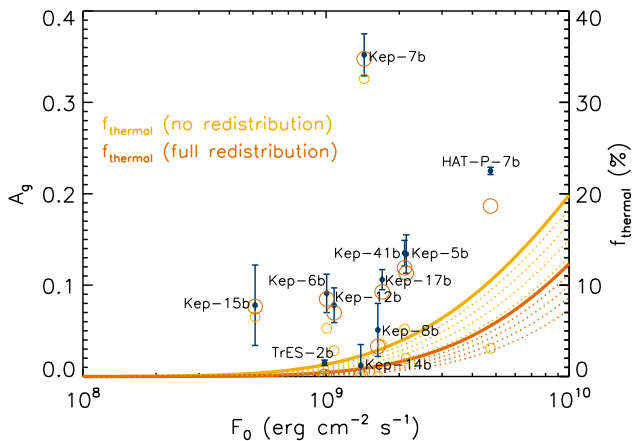}
\caption{Geometric albedo versus the incident stellar flux for a sample of hot Jupiters studied with \textit{Kepler} photometry.  Error bars include possible contamination by thermal emission.  The curves labeled ``$f_{\rm thermal}$" are estimations of the fraction of thermal flux radiated in the \textit{Kepler} bandpass, for $A_g=0$ (solid curve) and $A_g=0.1$, 0.2, 0.3 and 0.4 (dotted curves).  The small and large open circles represent the geometric albedos corrected for contamination by thermal emission (see text) assuming no and full redistribution, respectively.}
\label{fig:kepdata}
\end{figure}

Built primarily to study the occurrence of Earth-like exoplanets in our local cosmic neighborhood, the actual scientific scope of the \textit{Kepler Space Telescope} has expanded to include the study of hot Jupiters and their atmospheres.  The detection of transits and eclipses in the broadband optical channel of \textit{Kepler} has enabled the geometric albedos of about a dozen hot Jupiters to be measured.  In Figure \ref{fig:kepdata}, we show the geometric albedo ($A_g$) versus the incident stellar flux at the substellar point ($F_0$) for a sample of hot Jupiters, for which we have performed an improved analysis of Q0--Q14 data (see \S\ref{subsect:data} for details).  To begin our discussion requires defining a few temperatures \citep{ca11},
\begin{equation}
\begin{split}
&T_0 \equiv T_\star \left( \frac{R_\star}{a} \right)^{1/2}, \\
&T_{\rm eq} \equiv T_\star \left( \frac{R_\star f_{\rm dist}}{a} \right)^{1/2} \left( 1 - A_{\rm B} \right)^{1/4},\\
&T_{\rm eq,0} \equiv T_\star \left( \frac{R_\star}{2a} \right)^{1/2},\\
\end{split}
\end{equation}
with $T_\star$ denoting the stellar effective temperature, $R_\star$ the stellar radius and $a$ the orbital semi-major axis of the exoplanet.  The quantity $T_0$ may be regarded as the ``irradiation temperature" at the substellar point, such that $F_0 = \sigma_{\rm SB} T_0^4 = 4\sigma_{\rm SB} T^4_{\rm eq,0}$ is the incident stellar flux (or ``stellar constant") with $\sigma_{\rm SB}$ denoting the Stefan-Boltzmann constant.  The quantity $T_{\rm eq}$ is our conventional definition of the ``equilibrium temperature" and includes a dimensionless factor $f_{\rm dist}$ that describes the efficiency of heat redistribution from the dayside to the nightside hemisphere: we have $f_{\rm dist} = 2/3$ and $1/2$ for no and full redistribution, respectively \citep{hansen08}.  The quantity $T_{\rm eq,0}$ is the equilibrium temperature assuming full redistribution and a vanishing albedo.  Strictly speaking, the geometric albedo is defined at a specific wavelength $\lambda$ and at zero viewing angle.  The spherical albedo ($A_s$) is the geometric albedo considered over all viewing angles \citep{russell16,sudarsky00,seager10,mb12}.  For a Lambertian sphere (isotropic scattering), we have $A_g = 2 A_s / 3$.  When integrated over all wavelengths, we obtain the Bond albedo ($A_{\rm B}$).  Since we are dealing with observations integrated over a broad optical bandpass, we assume $A_{\rm B} = 3 A_g/ 2$; the consideration of more sophisticated scattering behavior will introduce an order-of-unity correction factor to this relation.

For the range of $T_{\rm eq,0}$ values listed in Table 1, we estimate that the hot Jupiters examined in Figure \ref{fig:kepdata} radiate mostly at wavelengths of about 1--2 $\mu$m (using Wien's law).  Nevertheless, since the \textit{Kepler} bandpass extends from $\lambda_1 \approx 0.4$ $\mu$m to $\lambda_2 \approx 0.9$ $\mu$m \citep{Koch:2010}, it is instructive to estimate the fraction of thermal flux from the exoplanet radiated in this bandpass,
\begin{equation}
f_{\rm thermal} = \frac{\pi \int^{\lambda_2}_{\lambda_1} B_\lambda\left(T_{\rm eq}\right) ~d \lambda}{\sigma_{\rm SB} T^4_{\rm eq}},
\end{equation}
by approximating the spectral energy distribution of a hot Jupiter to be a blackbody function.  In Figure \ref{fig:kepdata}, we plot $f_{\rm thermal}$ as a function of $F_0$ for both $f_{\rm dist}=2/3$ and $1/2$.  It is apparent that $f_{\rm thermal}$ may not be small and generally increases with $F_0$, implying that the measured geometric albedo \citep{seager10},
\begin{equation}
A_g=\frac{F_{p,\oplus}}{F_{\star,\oplus}}\left(\frac{a}{R_p}\right)^2,
\label{eq:ag}
\end{equation}
may be contaminated by thermal emission ``leaking" into the \textit{Kepler} bandpass, causing $A_g$ to be over-estimated.  The quantities $F_{p,\oplus}$ and $F_{\star,\oplus}$ are the fluxes from the star and the exoplanet, respectively, received at Earth, while the radius of the exoplanet is given by $R_p$.  One may approximately correct for the contamination by thermal emission by considering the following equation,
\begin{equation}
\begin{split}
A_g &= \left[ \frac{F_{p,\oplus}}{F_{\star,\oplus}} - \frac{\pi \int^{\lambda_2}_{\lambda_1} B_\lambda\left(T_{\rm eq}\right) ~d \lambda}{F_0} \left( \frac{R_p}{R_\star} \right)^2 \right] \left( \frac{a}{R_p} \right)^2 \\
&= A_{g,{\rm obs}} - \frac{\pi \int^{\lambda_2}_{\lambda_1} B_\lambda\left(T_{\rm eq}\right) ~d \lambda}{F_0} \left( \frac{a}{R_\star} \right)^2,
\end{split}
\end{equation}
where $A_{g,{\rm obs}}$ is the measured value of the geometric albedo obtained by applying equation (\ref{eq:ag}).  Since $T_{\rm eq}$ depends on $A_g$, the preceding expression is an implicit equation for the geometric albedo, which may be solved to obtain the ``de-contaminated" $A_g$, also shown in Figure \ref{fig:kepdata}.  The small and large open circles represent the corrected $A_g$ values assuming no and full redistribution, respectively.  Generally, decreasing the efficiency of heat redistribution decreases the geometric albedo obtained.  A better approach is to allow for $f_{\rm dist}$ to vary with $F_0$, since the efficiency of heat redistribution worsens as $F_0$ increases \citep{php12}, but we do not attempt this as the functional form of $f_{\rm dist}(F_0)$ is not well known.  For this reason, we do not specify the uncertainties associated with the corrected $A_g$ values.

Based on the results in Figure \ref{fig:kepdata}, there are two possible interpretations:
\begin{itemize}

\item Taken at face value (without performing a correction for contamination by thermal emission), the $A_g$ versus $F_0$ data exhibits a weak correlation (Spearman rank coefficient of 0.6), although the $A_g = 0.352 \pm 0.023$ measurement associated with Kepler-7b stands out.  When we correct for contamination by thermal emission assuming full redistribution ($f_{\rm dist}=1/2$), the $A_g$ values do not change much (and the Spearman rank coefficient remains, to the first significant figure, unchanged).

\item When the correction is performed assuming no redistribution ($f_{\rm dist}=2/3$), the trend flattens as expected (with a Spearman rank coefficient of $-0.1$).  Three of the data points are consistent with being zero.  The key point is that the correlation between $A_g$ and $F_0$ can only weaken, and not strengthen, when heat redistribution is taken into account.

\end{itemize}
Our conclusion is that there exists no clear correlation between $A_g$ and $F_0$.  Values of $A_g \approx 0.1$ may be consistent with Rayleigh scattering caused by hydrogen molecules alone \citep{sudarsky00}, without the need for the presence of clouds or condensates.  The high geometric albedo associated with Kepler-7b ($A_g \approx 0.35$) may require an explanation that includes the effects of clouds or condensates \citep{demory11,demory13}.

\subsection{Theoretical Motivation}

Unlike other previously examined relationships between various properties of hot Jupiters (e.g., radius and heat redistribution versus $T_{\rm eq}$; e.g., \citealt{ca11,ds11,laughlin11,php12}), there is no clear trend of $A_g$ with the incident stellar flux.  One of the goals of the present study is to suggest that the absence of a clear trend is caused by a combination of opacity effects, possibly due to the presence of condensates or clouds, and atmospheric circulation, the latter of which is often ignored in spectral analyses of hot Jupiters.  The study of clouds or hazes is emerging as a major theme in the observations of hot Jupiters (e.g., \citealt{lec08,pont08,sing11,gibson12}) and directly-imaged exoplanets (e.g., \citealt{barman11,madhu11,marley12,lee13}), and has long been an obstacle plaguing advances in the understanding of brown dwarfs (e.g., \citealt{sm08,artigau09,burrows11,helling11,buenzli12}).  

On the theoretical front, several trends are now understood:
\begin{itemize}

\item The strength and depth of atmospheric circulation is intimately tied to the intensity of stellar irradiation \citep{php12}.  An ``eddy diffusion coefficient" ($K_{\rm zz}$) is often used to \emph{mimic} this behavior, but ultimately the relationship between atmospheric circulation and stellar flux can---and should---be calculated from first principles using global, three-dimensional (3D) simulations.

\item The geometric albedo of an irradiated atmosphere controls not only the amount of starlight penetrating the atmosphere, but also the \emph{depth} of the penetration \citep{fortney08,hhps12,dd12}.

\item Whether a particle embedded in an atmospheric flow is held aloft depends not only on its size and density, but also on the local temperature, pressure and velocity field of the atmosphere (e.g., \citealt{spiegel09}).  The abundance of the particles relative to the atmospheric gas, their sizes and their composition in turn determine the geometric albedo, which controls the heating of the atmosphere.

\end{itemize}
To date, no model or simulation has succeeded in including all of these effects, which need to be studied in concert in order to gain a global perspective on the atmospheres of irradiated exoplanets.  Examining each factor in isolation is reasonable and necessary as an initial step, but the final word will come from studying their complex interplay.  Conversely, investing precision in some of these factors at the expense of others may yield misleading results.  We highlight some features of this interplay using an analytical toy model and scaling relations in the hope that it will motivate more sophisticated, follow-up studies.  More succinctly, the goal of the present study is to relate the presence of condensates and their properties with the geometric albedo, the pressure of the atmospheric layer probed by the \textit{Kepler} bandpass and the strength of atmospheric circulation, which is tied to the stellar irradiation flux.

To begin, \S\ref{sect:method} describes the details involved in constructing Figure \ref{fig:kepdata} as well as our theoretical method.  Our results are presented in \S\ref{sect:results}.  An example of comparative exoplanetology is described in \S\ref{sect:discussion} (Kepler-7b versus Kepler-12b), along with an ``observer's cookbook" for diagnosing qualitative trends associated with the geometric albedo and phase curves.

\section{Methodology}
\label{sect:method}

\begin{table*}
\label{tab:data}
\begin{center}
\caption{Geometric Albedos of Hot Jupiters from \textit{Kepler} Photometry, Refined Using Q0--Q14 Data}
\begin{tabular}{lcc}
\hline
\hline
Object Name & $A_g$ & $T_{\rm eq,0}$ (K)  \\
\hline
TrES-2b     &    $0.015 \pm  0.003$ &  $1444  \pm  13$ \\
HAT-P-7b    &    $0.225 \pm  0.004$ &  $2139  \pm  27$ \\
Kepler-5b   &    $0.134 \pm  0.021$ &  $1752  \pm  17$ \\
Kepler-6b   &    $0.091 \pm  0.021$ &  $1451  \pm  16$ \\
Kepler-7b   &    $0.352 \pm  0.023$ &  $1586  \pm  13$ \\
Kepler-8b   &    $0.051 \pm  0.029$ &  $1638  \pm  40$ \\
Kepler-12b  &    $0.078 \pm  0.019$ &  $1477  \pm  26$ \\
Kepler-14b  &    $0.012 \pm  0.023$ &  $1573  \pm  26$ \\
Kepler-15b  &    $0.078 \pm  0.044$ &  $1225  \pm  31$ \\
Kepler-17b  &    $0.106 \pm  0.011$ &  $1655  \pm  40$ \\
Kepler-41b  &    $0.135 \pm  0.014$ &  $1745  \pm  43$ \\
\hline
\end{tabular}
\end{center}
\end{table*}

\subsection{Data}
\label{subsect:data}

Our goal is to perform detailed photometric analyses of all {\it Kepler} giant exoplanets confirmed to date in order to precisely constrain their geometric albedos.  In this section, we describe our methodology and results.

\subsubsection{Kepler photometry}

This study is based on quarters Q0 through Q14 of {\it Kepler} data \citep[see][for the Q0--Q8 data release]{Batalha:2013} that are available at the time of writing. In total, the datasets encompass about 1,250 days of quasi-continuous photometric monitoring between May 2009 and October 2012. We retrieved the Q0--Q14 FITS files from MAST\footnote{http://archive.stsci.edu/kepler/} and extracted the raw long-cadence photometry \citep{Jenkins:2010a} for each target.

\subsubsection{Data analysis and derivation of the geometric albedo}

We consider the following 11 giant exoplanets in this study: TrES-2b \citep{Barclay:2012a}, HAT-P-7b \citep{Pal:2008,Christiansen:2010a}, Kepler-8b \citep{Jenkins:2010}, Kepler-6b \citep{Dunham:2010}, Kepler-5b \citep{Koch:2010}, Kepler-12b \citep{fortney11}, Kepler-7b \citep{latham10}, Kepler-14b \citep{Buchhave:2011}, Kepler-15b \citep{Endl:2011}, Kepler-41b \citep{Santerne:2011b,Quintana:2013} and Kepler-17b \citep{Desert:2011b}. The purpose of this analysis is to search for the planetary occultation (whose depth yields $F_{p,\oplus}/F_{\star,\oplus}$) in the {\it Kepler} bandpass and derive the corresponding geometric albedo.  To this end, we employ a Markov Chain Monte Carlo (MCMC) framework to characterize the posterior distribution of $A_g$. MCMC is a Bayesian inference method based on stochastic simulations that samples the posterior probability distributions of adjusted parameters for a given model. Our MCMC implementation \citep[described in, e.g.,][]{Gillon:2012} uses the Gibbs sampler and the Metropolis-Hastings algorithm to estimate the posterior distribution function of all jump parameters. Our nominal model is based on a star and a transiting planet on a Keplerian orbit about their center of mass.

Input data provided to the MCMC for each system consist of the Q0--Q14 {\it Kepler} photometry and the spectroscopic stellar parameters (effective temperature $T_\star$, metallicity [Fe/H] and $\log g_\star$) published in the references mentioned above. We correct for the photometric dilution induced by neighbor stellar sources using a quarter-dependent dilution factor based on the dilution values presented in the literature and on the contamination values reported in the FITS files headers \citep{Bryson:2010} . 

We divide the total lightcurve in segments of duration of about 24 to 48~hr and fit for each of them the smooth photometric variations due to stellar variability or instrumental systematic effects with a time-dependent quadratic polynomial. Baseline model coefficients are determined at each step of the MCMC for each lightcurve with the singular value decomposition method \citep{Press:1992}. The resulting coefficients are then used to correct the raw photometric lightcurves.

We assumed a quadratic law for the stellar limb-darkening (LD) and used $c_1=2u_1+u_2$ and $c_2=u_1-2u_2$ as jump parameters, where $u_1$ and $u_2$ are the quadratic coefficients. $u_1$ and $u_2$ were drawn from the theoretical tables of \citet{Claret:2011} for the stellar parameters obtained from the references above.

The MCMC has the following set of jump parameters: the planet/star flux ratio, the impact parameter $b$, the transit duration from first to fourth contact, the time of minimum light $t_0$, the orbital period, the occultation depth, the two LD combinations $c_1$ and $c_2$ and the two parameters $\sqrt{e}\cos\omega$ and $\sqrt{e}\sin\omega$. The latter two parameters allow to fit for the occultation phase and width. A uniform prior distribution is assumed for all jump parameters but $c_1$ and $c_2$, for which a normal prior distribution is used, based on theoretical tables and the stellar parameters used as input data to the MCMC fit.

We run two Markov chains of 100,000 steps each. The good mixing and convergence of the chains are assessed using the Gelman-Rubin statistic criterion \citep{Gelman:1992}. We use the posterior distribution functions for the jump parameters $F_{p,\oplus}/F_{\star,\oplus}$, $a/R_\star$ and $(R_p/R_{\star})^2$ obtained from the MCMC to derive the geometric albedo posteriors for each planet. We show the median of the posterior distribution function and its associated 1-$\sigma$ probability interval for $A_g$ and $T_{\rm eq,0}$ in Table~\ref{tab:data}.

As already mentioned in \S\ref{sect:intro}, the non-parametric Spearman rank coefficient is 0.6 for the $A_g$ versus $F_0$ data shown in Figure \ref{fig:kepdata}.  Values of $\pm 1$ indicate a perfectly monotonically increasing or decreasing trend, while those close to zero indicate no correlation between the two quantities.

\subsection{Models}

Any model constructed to describe clouds in exoplanetary atmospheres has to include both their radiative and dynamical effects.  Even when clouds contribute negligible mass to the atmosphere, they introduce strong radiative forcing to it in the form of the reflection of incident starlight (cooling) and the retention of reprocessed starlight in the infrared (heating).  On Earth, these two significant effects almost cancel, but not quite---it is this imperfect cancellation that is important for determining details of the terrestrial weather and climate system.  Getting these details correct to high precision remains a formidable challenge (see, e.g., \citealt{p10}).  Clouds also interact with the atmospheric flow.  To first order, we may neglect the dynamical effects of clouds on the flow (unless the dust-to-gas ratio is close to unity), but we may not neglect the dynamical effects of the flow on the clouds.  Atmospheric circulation sets the background state of velocity, temperature, pressure and density---and all of its dependent quantities---that allows us to determine if a cloud particle will remain at a certain location in the atmosphere.

To place the present study in context, we note that the 3D simulations of \cite{parmentier13} examine the effects of the atmospheric flow on embedded tracers that mimic the presence of cloud particles, but these tracers do not feed back on the flow in any way, both dynamically and radiatively.  By contrast, the 3D simulations of \cite{dda12} include an extra opacity source to describe Rayleigh scattering, but do not consider the interaction between the flow and embedded particles.

It is practically impossible to include all of these effects in an analytical model in any rigorous manner.  Instead, we subsume the absorptivity of the clouds into a general, infrared, absorption opacity that also describes the atmospheric gas.  The geometric albedo is a consequence of the scattering and absorption properties of the atmosphere, as we will discuss in \S\ref{subsect:albedo}.  While being mindful of this fact, we prescribe the geometric albedo as a free parameter when computing the analytical temperature-pressure profile.  To describe the action of the atmospheric flow on the cloud particles, we use an analytical expression for the local terminal velocity of a given cloud particle and approximate the vertical velocity of the flow to be a fixed fraction of the local sound speed.  Finally, we discuss the relationship between the properties of the cloud particles and the albedo.

\subsubsection{Radiative Forcing}

We use analytical models of the temperature-pressure profiles of hot Jupiters in the present study from \cite{hhps12}, where the cloudfree models of \cite{guillot10} were generalized to include a non-zero albedo.  Whether the albedo is due to Rayleigh scattering associated with molecules or scattering associated with condensates or dust grains is unspecified.  These models solve the one-dimensional (1D) radiative transfer equation in the plane-parallel, two-stream approximation using the method of moments, while the incoming stellar radiation is approximated as a collimated beam.  The free parameters involved are the irradiation temperature ($T_0$), the Bond albedo ($A_{\rm B}$), the absorption opacity in the infrared ($\kappa_{\rm IR}$) and the absorption opacity in the optical ($\kappa_{\rm O}$).  We set collision-induced absorption, associated with hydrogen molecules, to be the dominant opacity source at pressures of 0.1 bar and greater ($\epsilon=51$ for a bottom pressure of 10 bar).  There is an option to specify a purely-absorbing cloud deck of intermediate width, which we ignore for the purpose of simplicity and clarity.  

Starlight incident upon an atmosphere is predominantly absorbed at the following pressure level, known as the ``photon deposition layer" \citep{hhps12},
\begin{equation}
P_{\rm D} = \frac{0.63 g}{\kappa_{\rm O}} \left( \frac{1- A_{\rm B}}{1 + A_{\rm B}} \right),
\end{equation}
where $g$ denotes the surface gravity of the exoplanet.  In a purely reflecting atmosphere ($A_{\rm B}=1$), there is no penetration of starlight ($P_{\rm D} = 0$).  Note that this unique relationship between the photon deposition layer and the Bond albedo only exists in the limit of uniform, vertical populations of scatterers and absorbers in the 1D model atmosphere.  When the assumption of spatial uniformity is relaxed, this unique relationship is broken and it is now possible to specify cloud decks of varying spatial and optical thicknesses, located at different altitudes, that will produce the same Bond albedo.  While being aware of this possibility, we do not explore this complexity for several reasons: we are interested in elucidating trends, rather than making detailed predictions; to develop intuition, we confine ourselves to analytical models as much as possible, and it has been previously shown that such a generalization breaks the analytical nature of the model for the temperature-pressure profile \citep{hhps12}; the atmospheric data associated with hot Jupiters is not (yet) of a high enough quality to warrant such sophisticated investigations (unlike, for example, in the case of brown dwarfs).

\subsubsection{Effects of Atmospheric Flow on Condensates}

In this sub-section, we use simple scaling relations to elucidate the relationship between vertical, atmospheric flow and its ability to loft condensates.  The mean free path of molecules within an atmosphere is $L = m/\rho \sigma_{\rm m}$ where $m$ is the mean molecular mass, $\rho$ is the mass density and $\sigma_{\rm m} \sim 10^{-15}$ cm$^2$ is the cross section for inter-molecular interactions.  To describe the influence of the atmospheric flow on a particle embedded within it, we use the analytical formulae previously described in \cite{lw03} and \cite{spiegel09} (and references therein).  The terminal velocity of the (spherical) particle is
\begin{equation}
v_{\rm t} = \frac{2 {\cal C} r^2_c \rho_c g}{\rho \nu}.
\end{equation}
Its internal mass density is $\rho_c$, while its radius is $r_c$.  The kinematic viscosity of the atmospheric gas is $\nu \approx L c_s$ with $c_s$ being the local sound speed.  The quantity ${\cal C}$ is a correction factor accounting for the enhancement of the terminal velocity in rarefied media,
\begin{equation}
{\cal C} = 1 + N_k \left[ 1.256 + 0.4 \exp{\left(-\frac{1.1}{N_k}\right)} \right],
\end{equation}
and depends on the Knudsen number,
\begin{equation}
\begin{split}
N_k &= \frac{L}{r_c} = \frac{k_{\rm B} T}{P \sigma_{\rm m} r_c} \\
&\sim 10 \left( \frac{T}{1500 \mbox{ K}} \right) \left( \frac{P}{0.1 \mbox{ bar}} \frac{r_c}{1 ~\mu\mbox{m}} \right)^{-1},
\end{split}
\end{equation}
which in turn depends on the pressure level ($P$) and local temperature ($T$) of the atmosphere considered.  The Boltzmann constant is given by $k_{\rm B}$.  

We define a quantity $S \equiv v_z/v_{\rm t}$, which describes whether a particle is likely to be lofted by atmospheric circulation.  We approximate the vertical component of the velocity to be
\begin{equation}
v_z =  {\cal M}_z c_s = {\cal M}_z \left( \frac{\gamma k_{\rm B} T}{m} \right)^{1/2},
\label{eq:vz}
\end{equation}
where $\gamma = 7/5$ is the adiabatic gas index.  The vertical Mach number ${\cal M}_z$ is approximated to be constant, although we fully anticipate that it generally has the functional form ${\cal M}_z = {\cal M}_z(P, T)$.  It follows that
\begin{equation}
S \approx \frac{9 {\cal M}_z \gamma k_{\rm B} T}{2 {\cal C} \rho_c r^2_c \sigma_{\rm m} g}.
\label{eq:s}
\end{equation}
When $S \ge S_0$, atmospheric circulation keeps the particles lofted.  When $S < S_0$, the particles sink under the action of gravity.  We expect $S_0 \sim 1$.  When $N_k \ll 1$, we get ${\cal C}\approx1$.  In the limit of $N_k \gg 1$, we obtain
\begin{equation}
\begin{split}
S &\approx \frac{2.7 {\cal M}_z \gamma P}{\rho_c r_c g} \\
&\sim 1 ~\left(\frac{{\cal M}_z}{10^{-6}} \frac{\gamma}{7/5} \frac{P}{0.1 \mbox{ bar}} \right) \left( \frac{\rho_c}{3 \mbox{ g cm}^{-3}} \frac{r_c}{1 ~\mu\mbox{m}} \frac{g}{10^3 \mbox{ cm s}^{-2}} \right)^{-1}.
\end{split}
\label{eq:s2}
\end{equation}
We thus expect micron-sized particles to be lofted by atmospheric circulation.

\begin{figure}
\centering
\includegraphics[width=\columnwidth]{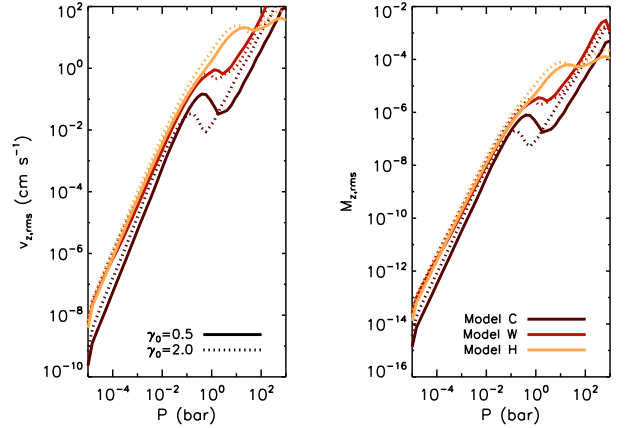}
\caption{Root-mean-square vertical velocity (left panel) and Mach number (right panel) as functions of pressure in the atmospheres of model hot Jupiters.  See text for details of the simulations.}
\label{fig:vert}
\end{figure}

To obtain the correct order-of-magnitude value for ${\cal M}_z$, we use the 3D simulations of atmospheric circulation of hot Jupiters by \cite{php12}.  Further details are described in \cite{hmp11} and \cite{hfp11}.  Specifically, we employ their Models C (``cold"), W (``warm") and H (``hot"), which have $T_{\rm eq,0} \approx 545$, 969 and 1723 K, respectively.  For each model, we also examine atmospheres with ($\gamma_0=2.0$) and without ($\gamma_0=0.5$) temperature inversions.  In Figure \ref{fig:vert}, we show the root-mean-square (rms) vertical velocity ($v_{z,{\rm rms}}$) and vertical Mach number (${\cal M}_{z,{\rm rms}}$) as functions of the vertical pressure $P$.  We see that at $P\sim 0.1$ bar, we have ${\cal M}_{z,{\rm rms}} \sim 10^{-6}$ with a somewhat weak dependence on the equilibrium temperature and the absence or presence of a temperature inversion.  A caveat is that the simulations of \cite{php12} essentially assume $A_g=0$ and do not include a treatment of scattering.  Secondly, like most other published general circulation models (GCMs) of hot Jupiters, they solve the primitive equations of meteorology, which assumes hydrostatic balance.  Hydrostatic balance does not preclude $v_z \ne 0$, since the vertical velocity is assumed to be sub-dominant only in the vertical component of the momentum equation, but it does imply that any simulated value is probably a lower limit to the one obtained by a fully non-hydrostatic simulation.  It is with these caveats that we adopt ${\cal M}_z = 10^{-6}$ at $P \sim 0.1$ bar.

Since we are interpreting $v_z$ to be the rms vertical velocity in equations (\ref{eq:vz}), (\ref{eq:s}) and (\ref{eq:s2}), $\min\{1,S\}$ may be interpreted as the relative abundance of embedded cloud particles summed horizontally over an entire atmospheric layer.

\subsubsection{Relationship Between Condensates and Albedo}
\label{subsect:albedo}

The scattering and absorption properties of cloud particles are mainly described by two quantities.  The first quantity is the ``single scattering albedo": $\omega_0 = \sigma_{\rm scat}/(\sigma_{\rm scat} + \sigma_{\rm abs})$ where $\sigma_{\rm scat}$ and $\sigma_{\rm abs}$ are the scattering and absorption cross sections, respectively.  The second quantity is the ``asymmetry parameter" ($g_0$), which is the mean cosine of the relative angle between the initial and scattered directions of a photon incident upon the particle.  It is the sole parameter in the Henyey-Greenstein scattering phase function, which determines the spatial distribution of scattered light \citep{hg41}.  Isotropic scattering occurs for $g_0=0$.  Particles described by $g_0=1$ produce predominantly forward scattering, while those described by $g_0=-1$ produce mostly backward scattering.   Generally, $\omega_0$ and $g_0$ depend on both the particle radius ($r_c$) and the wavelength considered ($\lambda$).  (See also \citealt{mb12}.)

Asymptotically, we expect that (e.g., \S5.4 of \citealt{p10})
\begin{equation}
\begin{cases}
g_0 \rightarrow 0, & r_c/\lambda \rightarrow 0, \\
g_0 \rightarrow G_0, & r_c/\lambda \rightarrow \infty,
\end{cases}
\end{equation}
where $G_0 > 0$ is a constant.  In other words, small particles scatter isotropically, while large ones tend to produce more forward scattering.  Whether a particle is ``small" or ``large" depends on the wavelength $\lambda$ considered---the relevant quantity is $2 \pi r_c/\lambda$, rather than $r_c$.  To compute the precise relationship between $r_c$, $g_0$ and $\lambda$ requires the full machinery of Mie scattering \citep{dl84,ld93,draine03}.  Generally, the scattering properties of a particle is determined mainly by its size and to a lesser extent by its composition, implying that a diagnosis of the cloud composition is a challenging and degenerate task \citep{p10}.

In an atmosphere populated by cloud particles, the geometric albedo $A_g$ is determined by both $\omega_0$ and $g_0$.  The simplest model one can construct of the function $A_g(\omega_0, g_0)$ involves solving the two-stream Schwarzschild equations with scattering, as described in \S5 of \cite{p10}.  Here, we generalize the formula to allow for $g_0 \ne 0$ and a finite blackbody efficiency of the cloud particle.  Denoting the incoming and outgoing fluxes by $F_\downarrow$ and $F_\uparrow$, respectively, we define
\begin{equation}
F_\pm \equiv F_\uparrow \pm F_\downarrow.
\end{equation}
Subtracting and adding the pair of Schwarzschild equations for $F_\uparrow$ and $F_\downarrow$ yields
\begin{equation}
\begin{split}
\frac{dF_-}{d\tau} =& -2 \epsilon_2 \left(1-\omega_0\right) F_+ + 4 \epsilon_2 \left(1-\omega_0\right) \pi B \\
&+ \omega_0 F_0 \exp{\left( \frac{\tau-\tau_0}{\cos \xi_0} \right)},\\
\frac{dF_+}{d\tau} =& -2 \epsilon_1\left(1-g_0\omega_0\right) F_- \\
&- 2\epsilon_1 \omega_0 g_0 \cos\xi_0 F_0 \exp{\left( \frac{\tau-\tau_0}{\cos \xi_0} \right)}.
\end{split}
\label{eq:twostream}
\end{equation}
Here, $\tau$ denotes the optical depth measured from some reference depth in the model atmosphere, while $\tau_0$ is the optical depth associated with the distance from this reference depth to the top of the model atmosphere.  The blackbody flux is given by $\pi B = \sigma_{\rm SB}T^4$.  The zenith angle $\xi_0$ is the angle between the incident stellar flux and the vertical axis; we allow for $\cos\xi_0 \ne 1$ in our derivation but later set it to be unity in our calculations.  The dimensionless quantities $\epsilon_1$ and $\epsilon_2$ are closure relations that depend on assumptions about the angular distribution of incoming versus outgoing radiation \citep{p10}.  Physically, the terms involving $F_0$ represent ``direct beam" emission from the star, while those involving $F_\pm$ represent the diffuse emission.

Taking the derivative of the second equation in (\ref{eq:twostream}) and eliminating $F_-$ using the first equation yields a second-order ordinary differential equation for $F_+$,
\begin{equation}
\begin{split}
&\frac{d^2 F_+}{d\tau^2} = K^2 F_+ - 8 \epsilon_1 \epsilon_2 \left( 1 - \omega_0 \right) \left( 1 - g_0 \omega_0 \right) \pi B\\
&- 2 \epsilon_1 \omega_0 F_0 \left[ 1 + g_0 \left( 1 - \omega_0 \right) \right] \exp{\left( \frac{\tau-\tau_0}{\cos \xi_0} \right)},
\end{split}
\label{eq:2ode}
\end{equation} 
where $K^2 \equiv 4 \epsilon_1 \epsilon_2 ( 1 - g_0 \omega_0 ) (1 - \omega_0)$.  The homogeneous solution for $F_+$ takes the form,
\begin{equation}
F_{+,{\rm h}} = f_\pm \exp{\left[ \pm K \left( \tau-\tau_0 \right) \right]}.
\end{equation}
For a deep atmosphere, we expect the homogeneous solution to not diverge when $\tau \rightarrow -\infty$, which compels us to set $f_-=0$.  The particular solution takes the form,
\begin{equation}
F_{+,{\rm p}} = f_{p,1} \exp{\left( \frac{\tau-\tau_0}{\cos \xi_0} \right)} + f_{p,2}.
\label{eq:particular}
\end{equation}
The constants $f_{p,1}$ and $f_{p,2}$ are determined by substituting equation (\ref{eq:particular}) into (\ref{eq:2ode}) and matching the coefficients found on both sides of the equation.  The full solution ($F_{+,{\rm h}}+F_{+,{\rm p}}$) then becomes
\begin{equation}
F_+ = f_+ \exp{\left[ K \left( \tau-\tau_0 \right) \right]} + f_{p,1} \exp{\left( \frac{\tau-\tau_0}{\cos \xi_0} \right)} + f_{p,2}.
\end{equation}
All that remains is to determine the constant $f_+$, which may be accomplished by substituting the full solution into the second equation in (\ref{eq:twostream}), which yields an expression for $F_-$ and hence $F_+ - F_- = 2 F_\downarrow$.  Applying the boundary condition of $F_\downarrow=0$ when $\tau=\tau_0$ produces an expression for $f_+$.

Returning to our expression for $F_-$, we again apply the condition $\tau=\tau_0$ to obtain $F_\uparrow = 3/2 A_g F_0 \cos\xi_0$, which yields the somewhat unwieldy formula for the geometric albedo,
\begin{equation}
\begin{split}
A_g =& \frac{2 \left [g_0 \omega_0 + \omega_0\left( 1 - g_0 \omega_0\right) \right]}{3\left( 1 - g_0 \omega_0 \right)\left( 1 - K^2 \cos^2\xi_0 \right)}\\
&+ \frac{K}{2 \epsilon_1 \left( 1 - g_0 \omega_0 \right) + K}\left[ \frac{16 \epsilon_1 \epsilon_2 f_{\rm B} \left( 1 - \omega_0 \right) \left( 1 - g_0 \omega_0\right)}{3 K^2} \right]\\
&- \frac{4 \epsilon_1 \left( 1 - g_0 \omega_0 \right)}{3 \left[2 \epsilon_1 \left( 1 - g_0 \omega_0 \right) + K\right]}\left( \frac{g_0 \omega_0}{1 - g_0 \omega_0} \right)\\
&- \frac{2 K}{3 \left[ 2 \epsilon_1 \left( 1 - g_0 \omega_0 \right) + K \right]}\\
&\times\frac{\left[1 + 2 \epsilon_1 \left( 1 - g_0 \omega_0 \right) \cos\xi_0 \right] \left[ g_0 \omega_0 + \omega_0 \left( 1 - g_0 \omega_0 \right) \right]}{\left( 1 - g_0 \omega_0 \right)\left( 1 - K^2 \cos^2\xi_0 \right)}.
\end{split}
\label{eq:formula}
\end{equation}
The preceding equation constitutes a solution of the radiative transfer equation.  The blackbody efficiency of the particle is defined as
\begin{equation}
f_{\rm B} \equiv \frac{\pi B}{F_0 \cos\xi_0}.
\end{equation}
In an isothermal atmosphere, we expect large particles to behave like blackbodies ($f_{\rm B}=1$) when they are observed at a wavelength $\lambda \ll 2\pi r_c$.  By contrast, small particles are inefficient emitters of radiation ($f_{\rm B}=0$) at a wavelength $\lambda \gg 2\pi r_c$.  

In the limit of $\omega_0=0$, we expect $3A_g/2 = f_{\rm B}$, which occurs only if $\epsilon_1 = \epsilon_2$.  We adopt the ``hemi-isotropic closure" ($\epsilon_1 = \epsilon_2=1$), which asserts that the flux is isotropic in each of the upward and downward hemispheres \citep{p10}.  We checked that equation (\ref{eq:formula}) yields $3A_g/2=1$ when $\omega_0=1$, independent of $g_0$ and $f_{\rm B}$.  In the limit of $g_0 = f_{\rm B}=0$, we recover equation (5.49) of \cite{p10},
\begin{equation}
\frac{3 A_g}{2} = \frac{\omega_0}{\left( 1 + 2\sqrt{1-\omega_0} \cos\xi_0\right)\left(1 + \sqrt{1-\omega_0}\right)}.
\end{equation}

\section{Results}
\label{sect:results}

\begin{figure}
\centering
\includegraphics[width=\columnwidth]{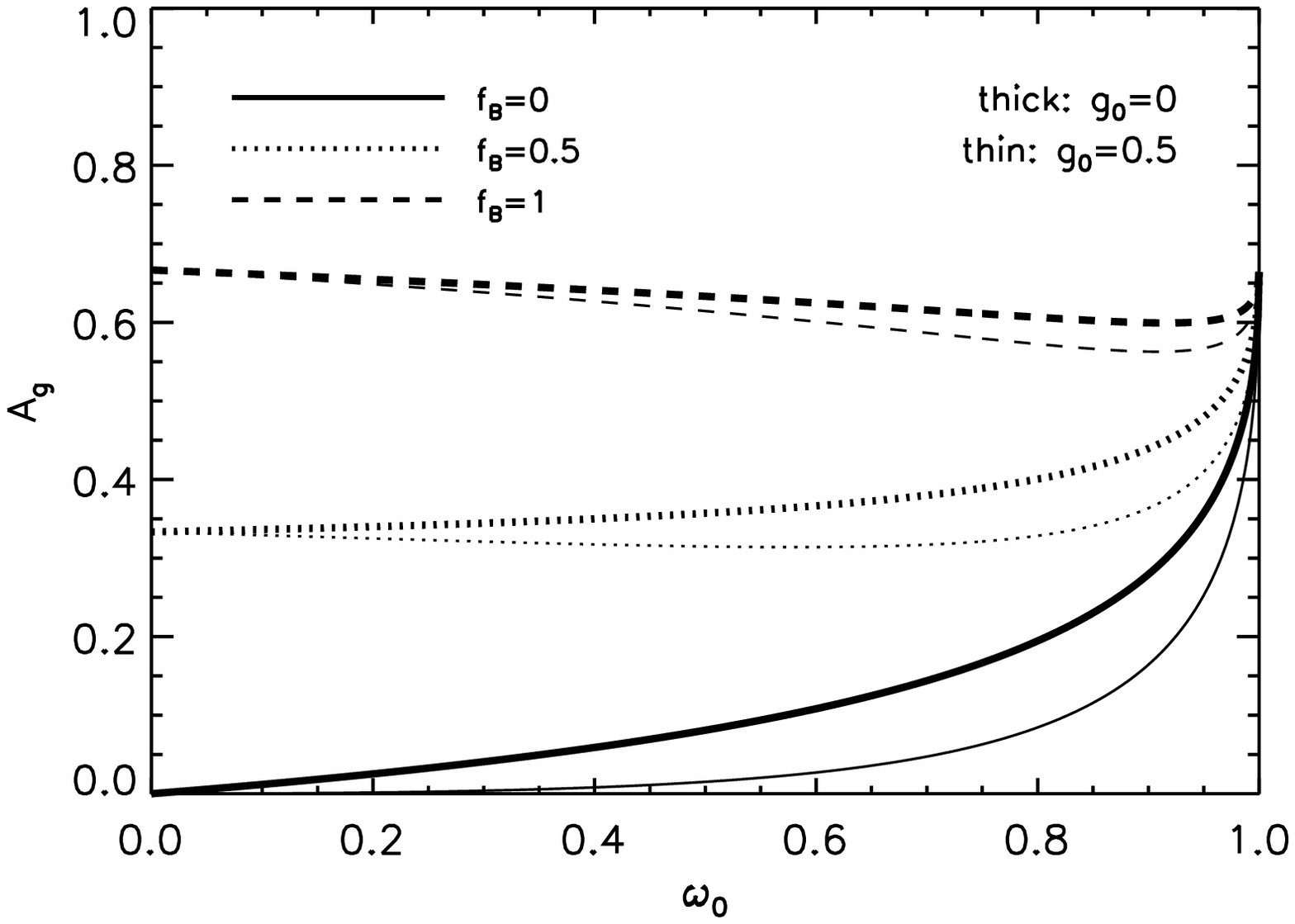}
\includegraphics[width=\columnwidth]{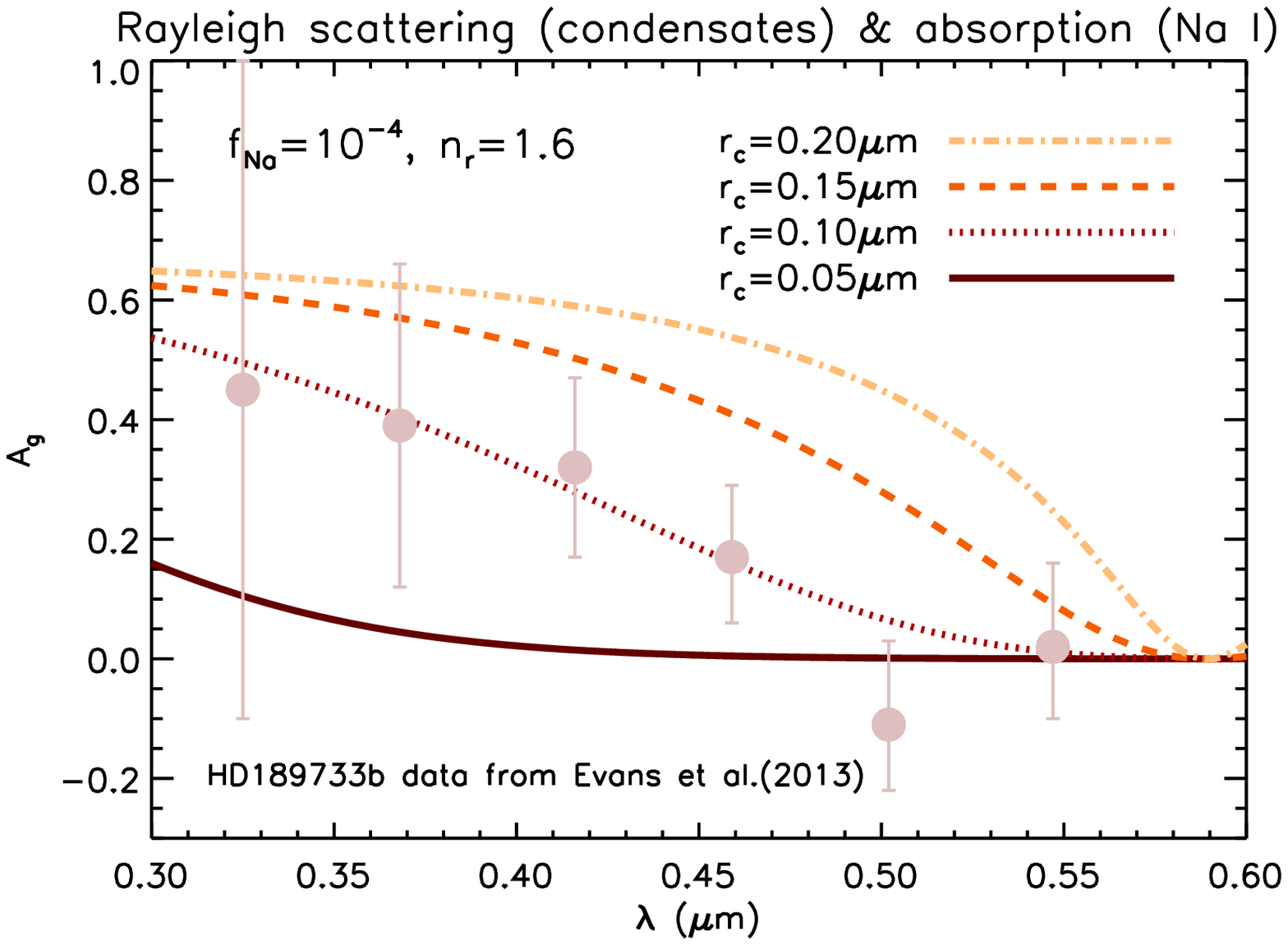}
\includegraphics[width=\columnwidth]{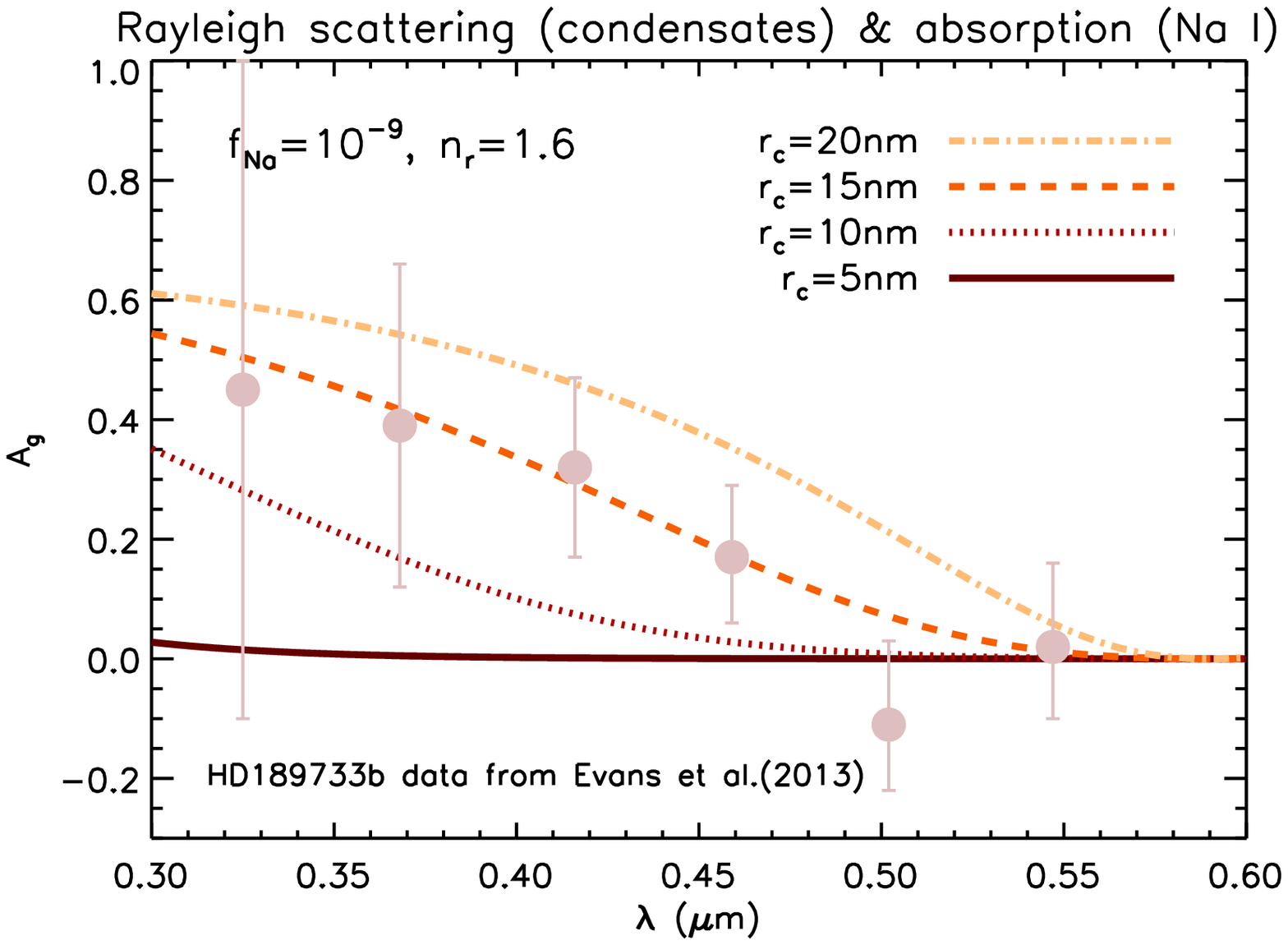}
\caption{Calculations of the geometric albedo.  Top panel: $A_g$ as a function of the single scattering albedo ($\omega_0$) for different values of the asymmetry parameter ($g_0$) and the blackbody efficiency of the particle ($f_{\rm B}$).  Middle panel: $A_g$ as a function of wavelength ($\lambda$) for Rayleigh scattering by condensates and absorption by the sodium D doublet assuming particle radii of $r_c \sim 0.1$ $\mu$m.  Bottom panel: same as for the middle panel, but for $r_c \sim 10$ nm.}
\label{fig:albedo}
\end{figure}

\begin{figure}
\centering
\includegraphics[width=\columnwidth]{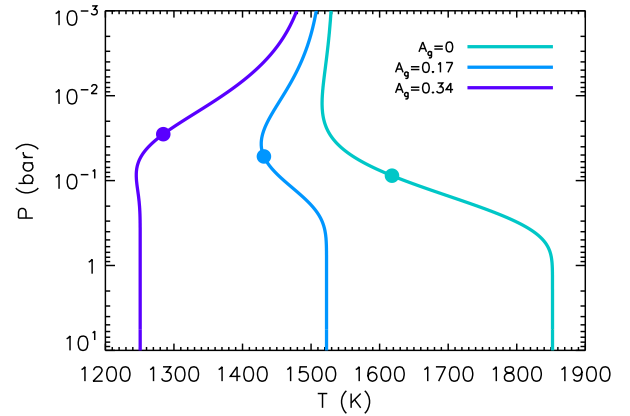}
\caption{Examples of temperature-pressure profiles adopting some of the parameters of Kepler-7b.  The dots indicate the locations of the photon deposition layers, where starlight is predominantly absorbed.  The \textit{Kepler} bandpass probes the photon deposition layer.}
\label{fig:tp}
\end{figure}

\subsection{Albedo Spectra: the Degeneracy Between Condensate Size and Relative Abundance to Sodium}

The top panel of Figure \ref{fig:albedo} shows $A_g$ as a function of $\omega_0$ for different values of $g_0$ and $f_{\rm B}$.  As the particle becomes more forward scattering ($g_0>0$), the geometric albedo generally decreases.  (See also \citealt{sudarsky00}.)  When $\omega_0 = 0$ and 1, $3A_g/2=f_{\rm B}$ and 1, respectively, as expected.  Generally, larger particles (higher $f_{\rm B}$ values) correspond to higher geometric albedos.  Next, we compute albedo spectra by considering the limiting case where Rayleigh scattering is entirely due to the presence of small condensates ($2\pi r_c/\lambda \ll 1$) and absorption is due to the sodium D doublet \citep{sudarsky00}.  (The potassium doublet absorbs at somewhat longer wavelengths: about 0.77 $\mu$m.)  Details of the scattering and absorption cross sections used are described in Appendix \ref{appendix}.  This simple model contains three parameters: the particle radius ($r_c$), the scattering refractive index of the condensates ($n_r$) and the relative abundance of sodium atoms to the condensates by number ($f_{\rm Na}$).  The single scattering albedo becomes
\begin{equation}
\omega_0 = \frac{\sigma_{\rm scat}}{\sigma_{\rm scat} + f_{\rm Na} \sigma_{\rm Na}},
\label{eq:rayleigh_sodium}
\end{equation}
and is somewhat insensitive to $n_r$.  We choose $n_r=1.6$ to mimic the presence of silicates such as enstatite.  We set $g_0=0$.  In the middle and bottom panels of Figure \ref{fig:albedo}, we show calculations of $A_g$ using equations (\ref{eq:formula}) and (\ref{eq:rayleigh_sodium}) for $r_c \sim 0.1$ $\mu$m and $\sim 10$ nm and compare them to the measurement of the albedo spectrum of the hot Jupiter HD 189733b by \cite{evans13}.  These assumed particle radii are consistent with those inferred by \cite{lec08}.  Generally, the computed geometric albedo is a degenerate function of $r_c$ and $f_{\rm Na}$.  However, since the Rayleigh cross section scales as $\sigma_{\rm scat} \propto r_c^6$, a small change in the particle radius needs to be compensated by a large change in the relative abundance of sodium atoms to condensates.  This property may prove to be useful for constraining $r_c$ in future observations of exoplanetary atmospheres.  We do not use equation (\ref{eq:formula}) to compute $A_g$ for use in the analytical temperature-pressure profiles, but rather specify it as a free parameter while being mindful that the geometric albedo is an emergent property of the scattering and absorption properties of the atmosphere.

\subsection{The Effects of Geometric Albedo on Thermal Structure}

Figure \ref{fig:tp} shows some examples of temperature-pressure profiles.  We have adopted some of the parameters of Kepler-7b: $T_{\rm eq} = 1586$ K and $\log{g} = 2.62$ \citep{demory11}.  We pick $\kappa_{\rm IR} = 0.004$ cm$^2$ g$^{-1}$ such that the infrared photosphere lies at $\sim 0.1$ bar.  Since the opacity sources in the optical range of wavelengths remain poorly known for hot Jupiters in general, we adopt $\kappa_{\rm O} = 0.003$ cm$^2$ g$^{-1}$ as an illustration.  We consider $A_g = 0, 0.17$ and 0.34 to demonstrate the effects of varying the albedo, corresponding to $P_{\rm D} \approx 0.09, 0.04$ and 0.03 bar.

In the limit of vertically uniform populations of scatterers and absorbers, starlight is absorbed and reflected mostly at the photon deposition layer ($P_{\rm D}$).  It is also the pressure level the \textit{Kepler} optical bandpass is probing.  Thus, the presence of a non-zero albedo has two effects.  The first, obvious one is to diminish the amount of heat deposited in an atmosphere.  The second, less obvious effect is to alter the \emph{location} at which most of the starlight is being deposited.  More reflecting atmospheres tend to have their starlight deposited higher in altitude.  Both properties are reflected in Figure \ref{fig:tp}.  Consequently, the effect of a non-zero albedo is to produce a temperature inversion if $A_g$ is of a high enough value.

\subsection{The Relationship Between Thermal Structure, Condensate Size and Condensate Abundance}

\begin{figure}
\centering
\includegraphics[width=\columnwidth]{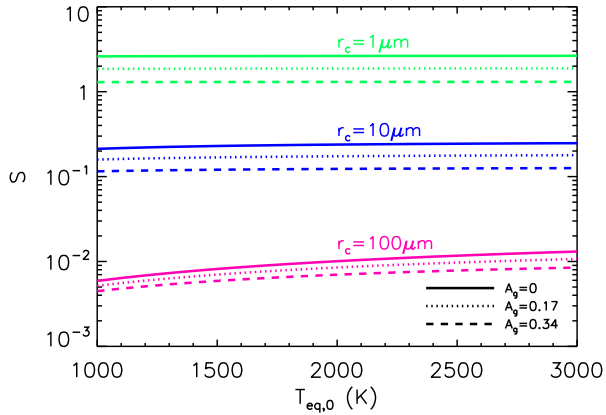}
\caption{The ``lofting parameter" $S$ as a function of $T_{\rm eq,0}$ (see text for definition) and the geometric albedo of the hot Jupiter for three different cloud particle radii ($r_c = 1$, 10 and 100 $\mu$m).}
\label{fig:sfactor}
\end{figure}

Next, we wish to examine the lofting properties of the atmosphere at $P=P_{\rm D}$.  We evaluate the temperature at the photon deposition layer using our analytical $T$-$P$ profiles: $T = T_{\rm D}$ where $T_{\rm D} \equiv T(P_{\rm D})$.  In Figure \ref{fig:sfactor}, we show $S$ as a function of $T_{\rm eq,0}$ for various values of $A_g$.  We show three sets of curves for $r_c = 1$, 10 and 100 $\mu$m.  For $r_c=1$ $\mu$m, we obtain $S \gtrsim 1$ for $T_{\rm eq,0}=1000$--3000 K, implying that micron-sized particles should be readily lofted in hot Jovian atmospheres, consistent with the results from the 3D simulations of \cite{parmentier13}.  Micron-sized particles have $N_k \gg 1$ and $S \propto P_{\rm D}$.  Since we do not allow the optical opacity ($\kappa_{\rm O}$) to depend on the intensity of incident, stellar irradiation, $S$ is a flat function of $T_{\rm eq,0}$.  However, since $P_{\rm D}$ depends on $A_g$, the pressure level probed is lower (higher altitude) and the corresponding value of $S$ is lower for larger values of $A_g$.  For $r_c = 100$ $\mu$m ($N_k \ll 1$), $S$ now has a dependence on $T_{\rm eq,0}$ as we have $S \propto T_{\rm D}$.  However, since $S \ll 1$ for all values of the equilibrium temperature examined, we do not expect particles with $r_c = 100$ $\mu$m to be lofted by atmospheric circulation.  Particles with $r_c = 10$ $\mu$m have $S \sim 0.1$--1 and are expected to be partially lofted as a population, again consistent with the results of \cite{parmentier13}.

If small cloud particles or grains ($N_k \gg 1$) are robustly formed in hot Jovian atmospheres, then they should be omnipresent due to the ease at which atmospheric circulation will keep them aloft.  The variations in their scattering and absorption properties (see \S\ref{subsect:albedo}), which is determined by their sizes and compositions, are expected to produce a scatter in the values of the measured geometric albedos.

\section{Discussion}
\label{sect:discussion}

\subsection{Comparative Exoplanetology: \\ Kepler-7b versus Kepler-12b}

The bounty of exoplanets discovered by the \textit{Kepler} mission has emphasized the importance of comparative exoplanetology.  Within our sample of hot Jupiters examined in Figure \ref{fig:kepdata}, Kepler-7b \citep{latham10,demory11} and Kepler-12b \citep{fortney11} provide for an intriguing comparison.  They receive similar degrees of stellar heating ($T_{\rm eq,0} \approx 1500$ K) and thus we expect the strength of atmospheric circulation to be comparable in both atmospheres.  They possess comparable surface gravities ($\log{g} \approx 2.6$) and radii ($R \approx 1.6$--$1.7 ~R_J$).  They orbit somewhat quiescent, Sun-like stars of comparable metallicity ([Fe/H]$\approx0.1$).  Yet their measured geometric albedos are non-negligibly different: $A_g \approx 0.35$ (for Kepler-7b) versus 0.08 (for Kepler-12b).  For Kepler-7b, there is evidence for the presence of condensates at the atmospheric layer probed by the \textit{Kepler} bandpass \citep{demory11,demory13}.  Furthermore, the optical phase curve of Kepler-7b exhibits a sinusoidal functional form with a peak that is offset from the secondary eclipse \citep{demory13}.  The infrared secondary eclipses of Kepler-7b, as detected by the \textit{Spitzer Space Telescope} at 3.6 and 4.5 $\mu$m, imply brightness temperatures that are markedly lower than that in the optical \citep{demory13}.  By contrast, the phase curve of Kepler-12b exhibits no sinusoidal variations (at the sensitivity level of \textit{Kepler}) with a period similar to the orbital motion of the exoplanet, while registering brightness temperatures of about 1400-1600 K in the infrared \citep{fortney11}.

We apply the lessons learnt from the analytical models presented in this study.  All else being equal, we expect the photon deposition layer to reside at a higher altitude (lower pressure) for Kepler-7b due to its higher albedo.  If large cloud particles ($N_k \ll 1$) are embedded in the atmosphere of Kepler-7b, then we expect $S \propto T_{\rm D}$ to describe the lofting property of its photon deposition layer.  Large particles generally produce a higher geometric albedo if $r_c \gg \lambda/2\pi$.  There are two possible configurations of $T_{\rm D}$ that will produce a sinusoidal functional form for $S$ with a peak that is offset from the secondary eclipse.  The first and simplest configuration is a shifted Heaviside function, i.e., $T_{\rm D}$ has two values, one in each hemisphere, but it is translated in longitude by some amount.  In this case, the corresponding brightness temperature in the \textit{Kepler} bandpass as a function of \emph{orbital phase} will be a sinusoidal function that has a peak offset \citep{ca08}.  The second configuration is for $T_{\rm D} \propto \sin(\phi \pm \phi_0)$, with $\phi$ being the longitude of the exoplanet and $\phi_0$ being a constant offset, which also produces a sinusoidal function for $S$.  In the right circumstances (i.e., $r_c \sim 10$ $\mu$m), the variation in temperature and pressure may cause $S$ to possess values ranging from 0.1 to 1.  Such a variation in $S$ produces a longitudinal variation in the abundance of lofted cloud particles, which in turn produces a longitudinal variation in the associated albedo and the flux of reflected starlight.  The observations of Kepler-7b \citep{demory13} are consistent with such a scenario.  By contrast, if small cloud particles ($N_k \gg 1$) are embedded in the atmosphere of Kepler-12b, we expect $S \propto P_{\rm D}$, which produces a flat phase curve for a given albedo value if the opacity in the optical range of wavelengths is roughly constant with longitude.  In other words, when small grains are present, we expect their abundance to be zonally uniform; if $g$, $\kappa_{\rm O}$ and $A_g$ are constant, then $P_{\rm D}$ and hence $S$ are constant across longitude.  Small particles are consistent with a lower albedo if $r_c \ll \lambda/2\pi$.

However, there is an important detail in the optical phase curve of Kepler-7b that is difficult to reconcile with our simple explanation.  The peak of the optical phase curve peaks \emph{after} secondary eclipse, implying that the corresponding brightness map peaks \emph{westwards} of the substellar point \citep{demory13}.  Irradiated atmospheres in the hot Jupiter regime are expected to possess temperature maps that peak \emph{eastwards} of the substellar point \citep{sp11}, a theoretical expectation that is corroborated by 3D simulations of atmospheric circulation.  That this property is independent of whether Newtonian cooling (e.g., \citealt{sg02,hmp11}), dual-band radiative transfer \citep{hfp11,rm12,php12} or multi-wavelength radiative transfer \citep{showman09} is utilized suggests that it is a robust outcome of the hot Jupiter regime.  If an infrared phase curve of Kepler-7b is measured with this property (eastward shift), then it is a ``smoking gun" for the presence of condensates---reflected light and thermal emission are not tracing each other, implying that the spatial distribution of the condensates is being modified by atmospheric dynamics.

\subsection{Caveats and Future Work}

Several caveats and unexplored aspects provide opportunities for future work.  We have not used an eddy diffusion coefficient ($K_{\rm zz}$) to mimic atmospheric circulation and instead approximated the vertical velocity as a fixed fraction of the local sound speed.  While such an approach includes the effect of varying the stellar irradiation, since it is tied to the temperature-pressure profile, it fails to capture the dependence of the \emph{depth} of atmospheric circulation on the incident stellar flux \citep{fortney08,dd12,php12}.  The vigor of atmospheric circulation in irradiated exoplanetary atmospheres means that if clouds form, they will be well-mixed from $\sim 1$ mbar all the way down to $\sim 10$ bar, depending on $F_0$ and $A_g$.  Performing 3D simulations of atmospheric circulation with a diversity of cloud configurations will inform 1D models on how to ``paint" clouds onto their $T$-$P$ profiles.

Another important interplay we have not explored concerns atmospheric chemistry.  Specifically, the strength and spectral distribution of the incident stellar flux modifies the absorption ($\kappa_{\rm O}$) and scattering ($A_g$) properties of both the condensates and the gas in the irradiated atmosphere.  To understand the lofting behavior of small cloud particles as a function of the incident stellar flux requires this interplay to be elucidated.

In recent years, retrieval models, originally developed for the modestly-irradiated planets/moons of the Solar System, have been employed to infer the temperature-pressure profiles and atmospheric chemistry/composition of hot Jupiters \citep{madhu09,lee12,line12}.  While being an important development in the study of exoplanetary atmospheres, these published works have so far been ``blue sky" and omitted the effects of clouds.  \cite{bs12} specify the albedo as a free parameter in their analysis, but only include one of its effects as suggested by their use of the \cite{guillot10} model: the diminution of the incident stellar flux, but not the modification of its vertical absorption profile.  They also allow for the cloud-top pressure to be a fitting parameter.  To illustrate the importance of including a non-zero albedo, consider the limiting case of $A_g=1$, in which case we expect the photon deposition layer to reside at the top of the atmosphere and for the infrared photosphere to be undefined (if interior heat from the irradiated exoplanet is negligible).  When the geometric albedo is just below unity, we expect the infrared photosphere(s) to be located just below the top of the atmosphere.  When $A_g=0$, the contribution functions are correctly computed by the published retrieval models.  Since we expect physical quantities to vary continuously, the contribution functions, at various infrared wavelengths, should shift to lower pressures as $A_g$ increases, an effect that needs to be included in the retrieval models.  Future work that includes a simple model of clouds, with a small number of free parameters, in the retrieval technique will set some empirical constraints on the cloud properties, although some degeneracy is anticipated.

While 3D simulations that treat only the dynamical or radiative interaction between the atmospheric flow and the cloud particles are insufficient for a full understanding of the effects of clouds in irradiated exoplanetary atmospheres, neither is a 1D analytical model that employs simple, approximate treatments for these effects, as presented in this study.  Nevertheless, both approaches drive us toward constructing falsifiable models that are realistic yet simple enough to be confronted by observations and feasibly included in future 3D simulations.

\subsection{An ``Observer's Cookbook"}

To provide an executive summary of the observational relevance of our study, we highlight a few scenarios and suggest possible (and possibly non-unique) interpretations.  When the peak amplitude of the phase curve is on the order of the occultation depth, then we term it to be ``sinusoidal"; if not, we term it to be ``flat".  When the angular offset of the peak of the phase curve from secondary eclipse is $\sim 1^\circ$ or close to zero, we term it to be ``small".  Phase offsets $\sim 10^\circ$ are ``large".

\begin{itemize}

\item \textbf{High albedo, sinusoidal optical phase curve:} Large cloud particles or dust grains ($\sim 10$ $\mu$m).  As discussed, a possible example is Kepler-7b.

\item \textbf{Low albedo, flat optical phase curve:} Small cloud particles ($\ll 1$ $\mu$m).  As discussed, a possible example is Kepler-12b.

\item \textbf{High albedo, small infrared phase offset:} A high albedo implies that the photon deposition layer resides higher in the atmosphere.  With most of the starlight being deposited at lower pressures, the infrared photosphere also lies at lower pressures.  With the atmosphere being more radiative (or less advective) at higher altitudes, a small phase offset in the infrared is expected \citep{sg02,ca11a,php12,heng12}.  Conversely, a larger infrared phase offset is expected for a low albedo \citep{fortney08}.

\item \textbf{Low albedo, small infrared phase offset:} For highly-irradiated hot Jupiters, the intense stellar flux trumps any effect due to opacity (e.g., albedo) and small infrared phase offsets are generally expected \citep{php12}.  Conversely, hot Jupiters with lower levels of incident irradiation are expected to possess large infrared phase offsets.

\end{itemize}

As an example, we consider the prototypical case of the hot Jupiter HD 189733b, for which large peak offsets have been measured in the infrared phase curves \citep{knutson07,knutson09}.  This is consistent with the low intensity of stellar irradiation impinging upon its atmosphere ($T_{\rm eq,0} \approx 1200$ K).  Transit observations in the ultraviolet and optical range of wavelengths reveal a spectral slope, punctuated by sodium lines, consistent with Rayleigh scattering by condensates present at the day-night terminators \citep{lec08,pont08,pont13,sing11,gibson12}.  That the infrared peak offsets are not small suggests that the albedo associated with the condensates is small, or even close to zero, at and near the peak of the stellar spectrum.  Based on a tentative comparison of several brightness temperature points with an analytical temperature-pressure profile, \cite{hhps12} estimate that the Bond albedo of HD 189733b is about 0.1.  The albedo spectrum of HD 189733b suggests a low to vanishing geometric albedo in the \textit{Kepler} bandpass \citep{evans13}.  Optical phase curves of HD 189733b will further constrain the properties of the condensates, including if they are small or large (as already described by the preceding scenarios).

Case studies that contradict these scenarios (e.g., \citealt{crossfield10}) hint at the possibility of missing physics or chemistry and will inspire novel ways of thinking about irradiated exoplanetary atmospheres.  Some of the degeneracies described may be broken by examining the colors of these atmospheres.

\vspace{0.1in}
\textit{KH acknowledges financial and/or logistical support from the University of Bern, the Swiss-based MERAC Foundation and the University of Z\"{u}rich.  We are grateful to Bruce Draine, Jaemin Lee, Michael Gillon and Nikku Madhusudhan for useful conversations.  KH benefited from discussions conducted at the Exeter-Oxford exoplanet workshop in April 2013 and at the PPVI conference in July 2013.  We thank the anonymous referee for constructive comments that improved the quality and clarity of the manuscript.}

\appendix

\section{Modeling sodium absorption and Rayleigh scattering}
\label{appendix}

The quantum mechanical properties of the sodium D doublet are well-known (e.g., \citealt{draine11}).  The absorption cross section is
\begin{equation}
\sigma_{\rm Na} = \frac{\pi e^2}{m_e c} ~f_{lu} \phi_\nu,
\end{equation}
where $e$ is the elementary unit of electric charge, $m_e$ is the mass of the electron, $c$ is the speed of light, $f_{lu}$ is the oscillator strength and $\phi_\nu$ is the dimensionless line profile function, described by a Lorentz profile,
\begin{equation}
\phi_\nu = \frac{4 A_{ul}}{16 \pi^2 \left( \nu - \nu_{ul}\right)^2 + A^2_{ul}},
\end{equation}
normalized such that $\int \phi_\nu ~d\nu = 1$ over all frequencies ($\nu$).  The frequency of the line transition is given by $\nu_{ul}$ with ``$u$" and ``$l$" denoting the upper and lower atomic levels, respectively.  The properties associated with the sodium doublet are
\begin{equation}
\begin{split}
&\lambda = 0.5891582 ~\mu\mbox{m}: ~f_{lu} = 0.641, g_u = 4, g_l = 2, A_{ul} \approx 6.159 \times 10^7 \mbox{ s}^{-1}, \\
&\lambda = 0.5897558 ~\mu\mbox{m}: ~f_{lu} = 0.320, g_u = g_l = 2, A_{ul} \approx 6.137 \times 10^7 \mbox{ s}^{-1}.\\
\end{split}
\end{equation}
The Einstein A-coefficients are computed using
\begin{equation}
A_{ul} = \frac{8 \pi^2 e^2 \nu^2_{ul} g_l f_{lu}}{m_e c^3 g_u},
\end{equation}
with $g_l$ and $g_u$ being the statistical weights of the lower and upper atomic levels, respectively.  Essentially, there are no free parameters involved in specifying $\sigma_{\rm Na}$.  The assumption employed here is that Doppler broadening of the line may be neglected, otherwise a Voigt profile has to be used in place of the Lorentz profile.

Rayleigh scattering by small particles is also a well-known phenomenon with a cross section given by
\begin{equation}
\sigma_{\rm scat} = \frac{2 \pi^5}{3} \left( \frac{n_r^2-1}{n_r^2+2} \right)^2 r_c^6 \lambda^{-4},
\end{equation}
where $n_r$ is the real part of the index of refraction.  For molecules, it is \citep{p10}
\begin{equation}
\sigma^\prime_{\rm scat} = \frac{32 \pi^3}{3} \left( \frac{n_r^\prime-1}{n} \right)^2 \lambda^{-4},
\end{equation}
where $n^\prime_r$ is the real part of the index of refraction for the molecular gas and $n$ is its number density.  For molecular gas, we have $( n^\prime_r - 1 ) \ll 1$; for refractory condensates, we have $( n_r - 1) \sim 0.1$.  Rayleigh scattering by molecules is weaker ($f_{\rm gas} \sigma_{\rm scat}^\prime/\sigma_{\rm scat} < 1$) as long as the particle radius exceeds a critical value,
\begin{equation}
r_c > \left[ \frac{4}{\pi n} \frac{\left( n_r^\prime -1 \right) \left( n_r^2 + 2 \right)}{\left( n_r^2 -1 \right)} \right]^{1/3} f^{1/6}_{\rm gas},
\end{equation}
where $f_{\rm gas}$ is the relative abundance of molecules to condensates by number.  For $P = 0.1$ bar and $T=1500$ K, we have $n \approx 5 \times 10^{17}$ cm$^{-3}$.  Using $n^\prime_r = 1.0001$ and $n_r = 1.6$, we obtain $0.9 f^{1/6}_{\rm gas}$ nm for the critical particle radius.  For comparison, we note that the Bohr radius is about 0.05 nm.  The relative weakness of Rayleigh scattering by hydrogen molecules may produce an albedo spectrum that is too low if sodium atoms are abundantly present.  Measuring the abundance of sodium relative to hydrogen remains challenging, even for HD 189733b \citep{huitson12}.


\label{lastpage}

\end{document}